\documentclass[%
superscriptaddress,
 prl,
rsi,%
amsmath,amssymb,
reprint,%
]{revtex4-1}
\DeclareUnicodeCharacter{0308}{}
\usepackage{graphicx}
\usepackage{dcolumn}
\usepackage{xcolor}
\usepackage{float}

\begin{document}
\author{Kalyani Chordiya}
\thanks{These authors contributed equally to this work}
\affiliation{ELI-ALPS, ELI-HU Non-Profit Ltd., Wolfgang Sandner utca 3., Szeged, H-6728, Hungary}
\affiliation{Institute of Physics, University of Szeged, Dóm tér 9, H-6720 Szeged, Hungary}
\author{Victor Despr{\'e}}
\thanks{These authors contributed equally to this work}
\affiliation{Theoretische Chemie, PCI, Universit{\"a}t Heidelberg, Im Neuenheimer Feld 229, D-69120 Heidelberg, Germany}
\author{Balázs Nagyillés}
\affiliation{ELI-ALPS, ELI-HU Non-Profit Ltd., Wolfgang Sandner utca 3., Szeged, H-6728, Hungary}
\affiliation{Institute of Physics, University of Szeged, Dóm tér 9, H-6720 Szeged, Hungary}
\author{Felix Zeller}
\affiliation{Theoretische Chemie, PCI, Universit{\"a}t Heidelberg, Im Neuenheimer Feld 229, D-69120 Heidelberg, Germany}
\author{Zsolt Diveki}
\affiliation{ELI-ALPS, ELI-HU Non-Profit Ltd., Wolfgang Sandner utca 3., Szeged, H-6728, Hungary}
\affiliation{Institute of Physics, University of Szeged, Dóm tér 9, H-6720 Szeged, Hungary}
\author{Alexander I. Kuleff}
\email{alexander.kuleff@pci.uni-heidelberg.de}
\affiliation{ELI-ALPS, ELI-HU Non-Profit Ltd., Wolfgang Sandner utca 3., Szeged, H-6728, Hungary}
\affiliation{Theoretische Chemie, PCI, Universit{\"a}t Heidelberg, Im Neuenheimer Feld 229, D-69120 Heidelberg, Germany}
\author{Mousumi U. Kahaly}
\email{Mousumi.UpadhyayKahaly@eli-alps.hu}
\affiliation{ELI-ALPS, ELI-HU Non-Profit Ltd., Wolfgang Sandner utca 3., Szeged, H-6728, Hungary}
\affiliation{Institute of Physics, University of Szeged, Dóm tér 9, H-6720 Szeged, Hungary}

\title{Photo-ionization Initiated Differential Ultrafast Charge Migration: Impact of Molecular Symmetries and Tautomeric Forms}

\date{\today}

\begin{abstract}
Photo-ionization induced ultrafast electron dynamics is considered as a precursor to the slower nuclear dynamics associated with molecular dissociation. Here, using \textit{ab initio} multielectron wave-packet propagation method, we study the overall many-electron dynamics, triggered by the ionization of outer-valence orbitals of different tautomers of a prototype molecule with more than one symmetry element.
From the time evolution of the initially created averaged hole density of each system, we identify distinctly different charge dynamics response in the tautomers. We observe that keto form shows charge migration direction away from the nitrogen bonded with tautomeric hydrogen, while in enol -- away from oxygen bonded to tautomeric hydrogen. Additionally, the dynamics following ionization of molecular orbitals of different symmetry reveal that a' orbitals show fast and highly delocalized charge in comparison to a" symmetry. These observations indicate why different tautomers respond differently to an XUV ionization, and might explain the subsequent different fragmentation pathways.
An experimental schematics allowing detection and reconstruction of such charge dynamics is also proposed. Although the present study uses a simple, prototypical bio-relevant molecule, it reveals the explicit role of molecular symmetry and tautomerism in the ionization-triggered charge migration that controls many ultrafast physical, chemical, and biological processes, making tautomeric forms a promising tool of molecular design for desired charge migration.
\end{abstract}

\maketitle

\section*{Introduction}
Understanding the photo-ionization induced response of molecules to XUV-UV radiation is crucial to unravel various chemical processes, like the photo-protection and photo-damage of RNAs, DNAs, and bio-molecules in general \cite{LAPO, ELSAYED, BROWN}. Photo-damage effects are observed as irreversible DNA/RNA modifications, resulting from ring-opening in nucleobases, fragmentation \cite{uf_260}, and initialization of chemical reactions \cite{JENA}. On the other hand, photo-protective effects are traced in reversible reactions such as photo-tautomerization \cite{Shukla2002, Nach2011}. In fact, many molecular systems in nature display multiple elements of symmetry and tautomeric forms, which impact their photochemical response. Gaining a molecular-level understanding of the physicochemical factors influencing the above photochemical processes has been a long-standing goal, which, apart from its fundamental importance, can lead to wide-ranging technological applications. Before detrimental excited-state reactions can take place after photo-excitation or ionization, the initial response is dominated by pure electron dynamics, with timescales typically sub- to a few femtoseconds. This ultrafast response to external perturbation represents a rearrangement of the electronic cloud which can dictate the relaxation pathways of the molecule. It can, for example, weaken certain bonds and thus prepare the system for fragmentation. Therefore, the ability to track and understand the electron dynamics down to sub-femtosecond timescales in the building blocks of biomolecules is important to address and eventually control photo-triggered processes including photosynthesis, photocatalysis, and photodamage of biomolecules amongst many others. 
Many research efforts have recently been dedicated to study the ultrafast response of the electronic cloud of molecular systems triggered by photo-ionization and subsequent dynamics \cite{belshaw2012,calegari2014,marciniak2019electron,li2016photodissociation}. 
M\aa{}nsson \textit{et al.}~\cite{C7CP02803B} provided the first experimental insight into the mechanisms that govern the dynamics following the action of ionizing radiation on DNA bases, while Nachtigallovà \textit{et al.}~\cite{B806323K} explored their electronic interactions and electronically excited states. Time-resolved photoelectron imaging has been widely used on molecules such as adenine~\cite{doi:10.1021/jz500264c} and iodide-uracil complex~\cite{li2016photodissociation} to study the charge dynamics and underlying DNA damage mechanisms at molecular level. However, the pure electron dynamics immediately following the sudden photo-ionization has still not been investigated experimentally. Due to the rapid progress in attoscience \cite{calegari2016}, directly following the first instances after ionization is already within reach. Theoretical support is, however, needed to guide and interpret such ambitious experiments and with the present work we would like to make a step in this direction. 

\begin{figure}
    \centering
    \includegraphics[scale = 0.19]{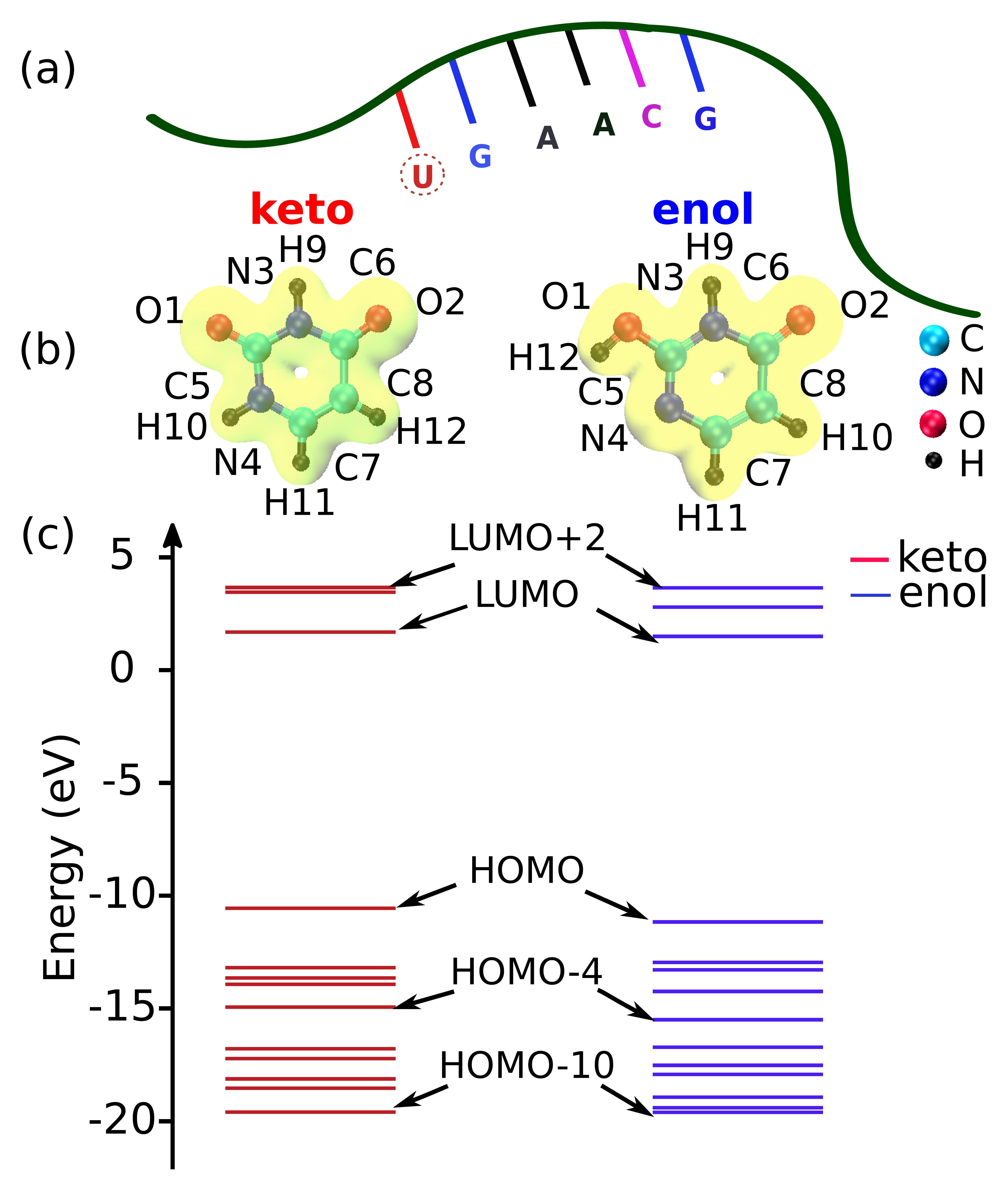}
    \caption{(a) Schematic of RNA strand with Uracil (U), along with other nucleobases, (b) Molecular structures of keto and enol and corresponding charge densities isosurfaces (0.03 e/\AA$^{3}$) (c) Molecular orbital energy levels.}
    \label{fig:introduction}
    \label{fig:spec}
\end{figure}

We choose a bio-relevant prototype molecule -- Uracil (\textbf{U}), which is found in multiple tautomeric forms  \cite{rejnek2005correlated}; the two most stable being 
`keto' and `enol' (with energy difference of 0.49~eV). Our choice of this molecule is further motivated by its omnipresence as nucleobase in biological macromolecules (see the schematic in Fig.~\ref{fig:introduction}(a)). 
It is important to stress that given the complexity of biological systems and the importance of the environment in their functionality, gas-phase analysis may not suffice for a complete understanding of the response of biomolecules to XUV irradiation. However, analyzing ultrafast charge migration at a building-block molecular level can provide a clearer picture for the basic mechanisms underlying the photo-induced response in bio-matter.

The tautomeric `keto' form with hydrogen on N4 (with two carboxyl groups) and `enol' form with H on O1 (with one carboxyl group and one hydroxyl group) are shown in Fig.~\ref{fig:introduction}(b). In 2018, Colasurdo \textit{et al.}~\cite{ue_tautomer} reported the coexistence of keto and enol tautomers of \textbf{U} and an intercoversion between the two forms at structural equilibrium \textbf{U} (see Fig.~\ref{fig:introduction} (b)). 
A study of XUV-induced structural changes responsible for the RNA breaking and associated hydrogen migration in 5-halouracil molecule within $>10$~fs was reported by Castrovilli \textit{et al.}~\cite{Castrovilli_2015}. Hence, \textbf{U} being a small and symmetric molecular structure with two stable tautomeric forms poses as an ideal prototype for this study.  

In this work, we study the ultrafast multi-electron dynamics of \textbf{U} after XUV photo-ionization. In the first few femtoseconds, the initial electronic response is dominated by the so-called correlation-driven charge migration \cite{CEDERBAUM1999205,kuleff2014ultrafast}. We focus on the analysis of this response in the different tautomeric forms to better understand how small structural modifications can actually impact this electron-correlation based mechanism. We investigate the global charge dynamics of the Uracil molecule, to understand the response of different tautomers after XUV photo-ionization. We also study specific correlation-driven charge migration dynamics initiated by ionization of individual molecular orbitals to elucidate whether tautomeric forms are a promising path for the development of an appropriate molecular design for a desired charge migration \cite{despre2019size}. 
Thus, our study serves as general proposition to identify the role of molecular symmetry and tautomerism in determining charge migration pathways and can be utilized to understand their possible response, for example, resistance to radiation damage in general. 

\section*{Computational methods and Theory}
\label{sect:Theory}

\begin{figure*}
\centering
\includegraphics[scale=0.6]{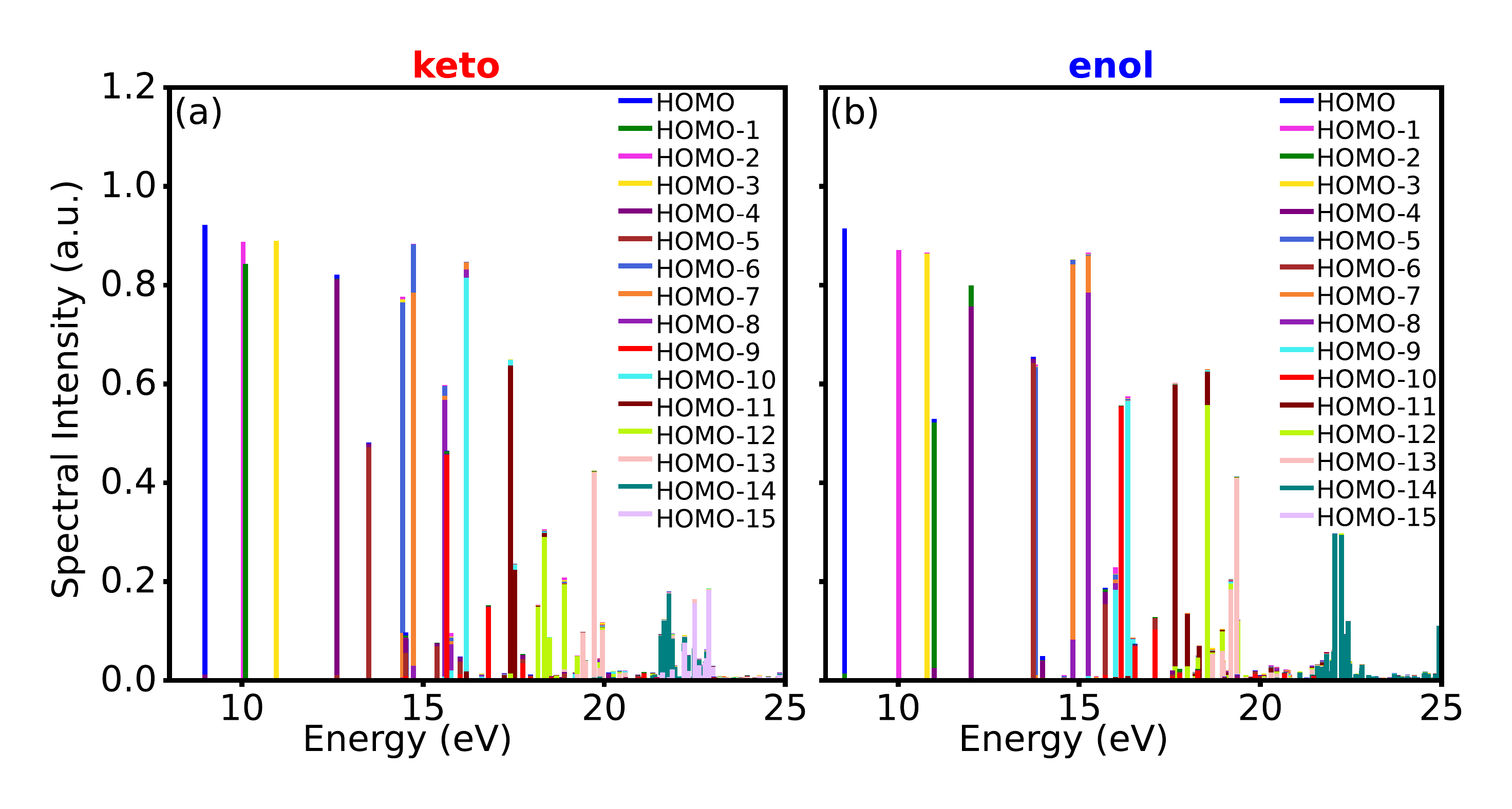}
\caption{Valence ionization spectra of (a) keto and (b) enol, computed with Green's function ADC(3) method.}
\label{fig:full_ionization}
\end{figure*}

The geometry optimization for the keto and enol tautomers is performed with density functional theory (DFT) using Gamess-UK package \cite{guest2005gamess} at the PBE0/def2-TZVP level \cite{adamo1999toward,B515623H}. The resulting bond lengths of the two molecules are tabulated in the supplementary information (\textbf{ESI$^\dag$}), Table~\ref{tab:Bond lengths}, and are in excellent agreement with the experimental values \cite{harsanyi1986equilibrium}. This suggests that the choice of basis set is appropriate for further calculations. Therefore, the same basis set was used to compute the Hartree-Fock (HF) orbital energies of the two tautomers, which are shown in Fig.~\ref{fig:introduction} (c). We see that the two molecules show different spacings between their orbital energies. A clear difference is observed, for example, for HOMO and HOMO$-1$ of keto (red) and enol (blue), or LUMO and LUMO$+1$. This shows that already at the HF level, i.e. without taking the electron correlation into account, the tautomeric forms possess different electronic structure. 
Considering this, different dynamics can be expected after ionization for both tautomers. 
In the following we will focus on dynamics occurring after photo-ionization of outer valence orbitals.

The formalism employed for tracing the ultrafast electron dynamics triggered by ionization has been used in numerous studies and has been described in detail elsewhere (see, e.g, \cite{kuleff2014ultrafast,Kuleff2018Ultrafast}). Here we will only briefly outline the key elements. The ionization spectra of the molecules are computed with the help of the non-Dyson ADC(3) scheme \cite{PhysRevA.28.1237, doi:10.1063/1.477085} for constructing the one-particle Green's function. The non-Dyson ADC scheme uses the so-called intermediate state representation of the secular matrix, yielding the construction of a complete basis through successive Gram-Schmidt orthogonalization of different classes of correlated excitations. The latter are built using the HF orbitals of the neutral ground state as a one-particle basis. The evolution of the initially created cationic state is computed with the help of the multielectron wave-packet propagation method \cite{kuleff2005multielectron}, based on a direct propagation with the non-Dyson ADC Hamiltonian using short-iterative Lanczos algorithm \cite{Lanczos}.

To trace the charge migration within the molecule and the associated electron dynamics, it is convenient to construct and analyze the time-dependent hole density $Q(\vec{r}, t)$  \cite{CEDERBAUM1999205, doi:10.1063/1.1540618}. As the name suggests, $Q( \vec{r}, t)$ describes the density of the hole at position $\vec{r}$ and time $t$ and is defined as the difference between the electronic density of the neutral molecule, described by its ground state $|\Psi_0\rangle$, and the electronic density of the evolving cation, described by the electronic wave packet $|\Phi_i(t) \rangle=e^{-i\hat{H}t} |\Phi_i(0)\rangle$
\begin{equation*}
Q(\vec{r},t) = \langle\Psi_{0}|\hat{\rho}|\Psi_{0}\rangle - \langle\Phi_{i}(t)|\hat{\rho}|\Phi_{i}(t)\rangle
\end{equation*}
\begin{equation}
\label{eq:hole denisty}
= \rho_{0}(\vec{r}) - \rho_{i}(\vec{r}, t).
\end{equation}
The density operator $\hat{\rho}$ in the second-quantization representation, using the HF orbitals $\phi_{p}( \vec{r})$ as one-particle basis, is given by
\begin{equation}
\hat\rho( \vec{r})= \sum_{p,q}\phi_{p}^{*}( \vec{r})\phi_{q}( \vec{r})\hat{a}^{\dagger}_{p}\hat{a}_{q},
\label{eq:1p_quantization}
\end{equation}
where the operator $\hat{a}^{\dagger}_{p}$ creates an electron in orbital $\phi_{p}$ and $\hat{a}_{q}$ destroys an electron to create a hole in orbital $\phi_{q}$. Within this representation, the hole density takes the form
\begin{equation}
    Q( \vec{r},t) = \sum_{p,q} \phi^{*}_{p}( \vec{r})\phi_{q}( \vec{r})N_{pq}(t),
\label{eq:q(r,t)_final}
\end{equation}
where $N_{pq}(t)$ is the so-called hole-density matrix. The diagonalization of the latter at each time point $t$ leads to the following form of the hole density 
\begin{equation}
    Q( \vec{r},t) = \sum_{p} |\tilde{\varphi}_{p}( \vec{r},t)|^2 \tilde{n}_{p}(t),
 \label{eq:q(r,t)}
\end{equation}
where $\tilde{\varphi}_{p}( \vec{r},t)$ are the natural charge orbitals and $\tilde{n}_{p}(t)$ their hole-occupation numbers. Further details on construction of the hole density employing the ADC approach can be found in Refs.~\cite{doi:10.1063/1.1540618, doi:10.1063/1.2428292, kuleff2005multielectron}.

\section*{Results and discussion}
\label{sec:uracil_keto and enol}

\begin{figure*}
    \centering
    \includegraphics[scale=0.19]{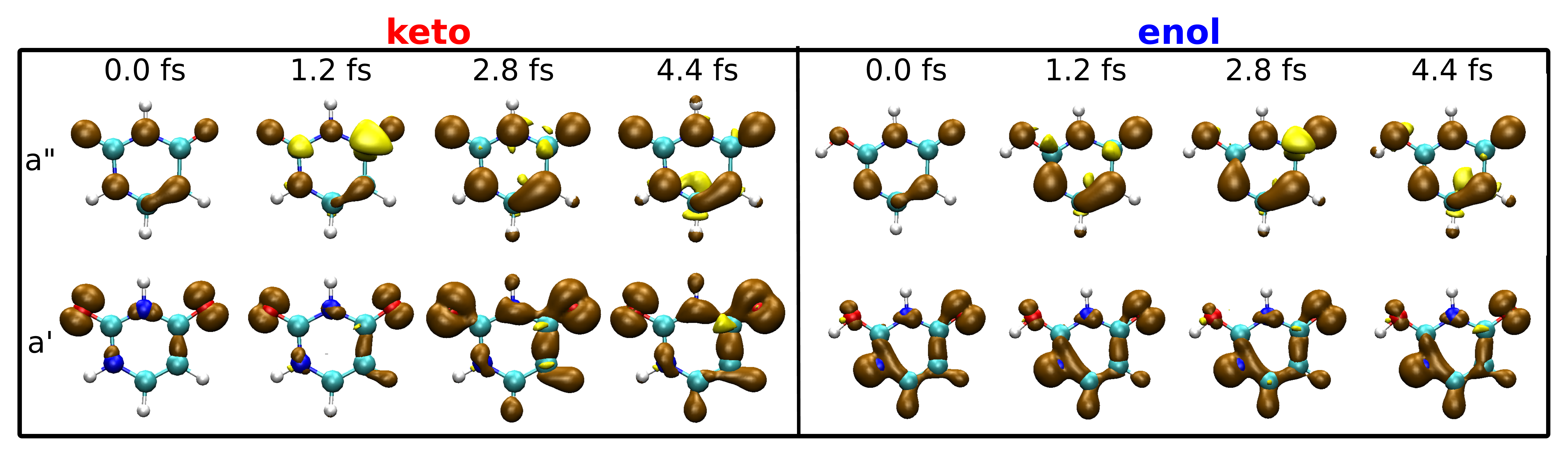}
    \caption{Snapshots of the evolution of the averaged hole density, Eq.~(\ref{eq:hole denisty}), following outer-valence ionization from molecular orbitals belonging to a" (top row) and a' (bottom row) symmetry of keto (left panel) and enol (right panel) at 0.0 fs, 1.2 fs, 2.8 fs, and 4.4 fs. The hole density (isosurface: 0.005 a.u.) is depicted in brown and the electron density (isosurface: -1E-6 a.u.), or the regions with excess of electrons, in yellow.}
    \label{fig:a' and a" mo ionization}
\end{figure*}

Let us first examine the ionization spectra of the chosen molecules. Both keto and enol tautomers of \textbf{U} are found in C$_s$ symmetry and have two irreducible representations: symmetric (a') and asymmetric (a"). Considering above-core twenty one occupied orbitals, HOMO$-20$ (9a') to HOMO (5a") of keto and enol, we find five orbitals in a" symmetry and the rest in a' symmetry. For keto a" orbitals are: HOMO (5a"), HOMO$-1$ (4a"), HOMO$-4$ (3a"), HOMO$-5$ (2a") and HOMO$-9$ (1a") and for enol are: HOMO (5a"), HOMO$-2$ (4a"), HOMO$-4$ (3a"), HOMO$-6$ (2a") and HOMO$-10$ (1a"). The ionization spectra up to 25~eV are shown in Fig.~\ref{fig:full_ionization}(a) for keto and \ref{fig:full_ionization}(b) for enol. Each line in these spectra represents a cationic eigenstate with a position corresponding to its ionization energy, and spectral intensity reflecting its ionization cross-section. The spectral intensity is given by the sum of the weights of all one-hole configurations (an electron removed from a single orbital) participating in the configuration-interaction expansion of the corresponding state. The weight of each one-hole configuration is color-coded in Fig.~\ref{fig:full_ionization}.
The first ionization energies (leftmost lines in Fig.~\ref{fig:full_ionization}) lie at 8.98~eV for keto (Fig.~\ref{fig:full_ionization}(a)) and 8.52~eV for enol (Fig.~\ref{fig:full_ionization}(b)) and are in close agreement with previously reported values: 9.20~eV (from experiment) \cite{yu1981ultraviolet} and 8.91~eV (obtained by G4, a high-level \textit{ab initio} method) \cite{doi:10.1002/jcc.24054}, respectively.

Significant differences between the ionization spectra of keto and enol forms are apparent. Indeed, we observe strong correlation effects in keto after 12.6~eV and in enol already from 10.8~eV on (see, Fig.~\ref{fig:full_ionization} (a) and (b), respectively). The lines below these energies are to a large extent pure one-hole (1h) states, resulting from the removal of an electron from a single orbital. This is the weak-correlation regime, where the molecular orbital picture still holds, and Hartree-Fork and Koopmans approximations \cite{szabo2012modern} are still meaningful. Above these energies (up to about 19~eV in Fig.~\ref{fig:full_ionization}(a) and 20~eV in Fig.~\ref{fig:full_ionization}(b)), with increasing strength of the electron correlation, the heights of the lines decrease, and weaker lines besides the main ones appear as satellite states. This is the strong-correlation regime, where the states have mostly two-hole--one-particle (2h1p) character, and thus describe additional excitations on top of the ionization. For deep-lying states closer to the core (see lines above 19~eV in Fig.~\ref{fig:full_ionization}(a) and 20~eV in Fig.~\ref{fig:full_ionization}(b)) no clear main lines can be distinguished, implying a complete breakdown of the molecular-orbital picture \cite{cederbaum1986correlation}. This is the regime of very strong correlation entering in the double-ionization continua of the molecules.

To shed some light on the charge dynamics initiated by ionization, we studied the global response of the two tautomers on an XUV photo-ionization that populates states in the weak- and strong-correlation regime. To do that, we computed the average electron dynamics, resulting from the single ionization of molecular orbitals lying up to about 15~eV, i.e. from HOMO to HOMO$-6$. 
This is a realistic situation, as the currently used attosecond XUV pulses, produced via high-harmonic generation process in rare gases, have typically spectra between 15 and 40 eV. These are also the characteristics currently available at ELI-ALPS High Repetition beamline \cite{ye2020attosecond}. Such pulses will be able to ionize all orbitals lying lower in energy, and thus populate all symmetry allowed cationic states below 15~eV. This energy was chosen also because above about 15~eV, the so-called correlation-band structure \cite{deleuze1996formation} starts to form, where the dynamics become strongly non-adiabatic, as demonstrated recently \cite{Herve2020Ultrafast}, which will necessitate a different theoretical approach.

We have to note that depending on the their spectral width, the attosecond XUV pulses will coherently populate different parts of the ionization spectrum of the molecules. To simulate the dynamics triggered by such pulses, one needs to coherently add the dynamics initiated by the removal of an electron from several orbitals. Many of those, as well as their weights, will strongly depend on the particular pulse characteristics used in experiment. As we are not aiming here at simulating a particular experiment, we prefer to study the averaged electron dynamics following outer-valence ionization, which better reflect the processes taking place in nature, where the ionization is performed with a non-coherent light. In this case, the hole density evolution is an incoherent sum of the charge migrations resulting from the removal of an electron from each of the energetically accessible orbitals.

Taking advantage of the possibility to align the molecule \cite{trabattoni2020setting,thesing2020effect}, a' or a'' symmetries of \textbf{U} can be studied separately by the linearly polarised light. We thus traced all correlation-driven charge migration processes initiated by removing electrons from HOMO to HOMO$-6$, by computing the evolution of the resulting hole density (see, Eq.~(\ref{eq:q(r,t)_final})). For each symmetry, the individual hole-density evolutions were then added and renormalized. The results are shown in Fig.~\ref{fig:a' and a" mo ionization}. The dynamical response of the systems to ionization from individual orbitals is presented and discussed in the \textbf{ESI$^\dag$}. 
Before we proceed with our analysis, let us touch upon the role of nuclear dynamics and its coupling to the electronic motion. In the present calculations, no non-adiabatic effects are included. This is a reasonable approximation if the created electron coherence lasts long enough, or at least longer than the electron dynamics process we are interested in. Previous works have reported that for correlation-driven charge migration (ionization of a single molecular orbital), a coherence longer than 10~fs can be expected \cite{despre2015attosecond, despre2018charge}. On the contrary, in the case where only weak correlation effects are present and the electronic wave packet has to be created by ionization out of 2 or more molecular orbitals, ultrafast decoherence due to the initial width of the wave packet (just a few fs) can take place \cite{vacher2017electron, arnold2017electronic}. This effect has even been shown at short time scales to be stronger for some systems than the impact of the nuclear motion itself \cite{vacher2016electron}. As we will see, the electron dynamics effects discussed in the present study mostly take place within the first 2~fs and thus the clamped-nuclei approximation used is justified. We note, however, that to fully understand the influence of the charge migration on the follow-up structural rearrangements in the uracil tautomers, the inclusion of the nuclear dynamics is certainly necessary.

The global response of uracil after ionization of a'' orbitals of the keto form, comparing the charge density at $t=0$~fs and $t=1.2$~fs  Fig.~\ref{fig:a' and a" mo ionization} (left panel, upper row), shows an overall decrease in hole density (brown iso-surface) on O1, O2, N3, N4, C7, and C8 and an increase in electron density (yellow iso-surface) on C5 and C6. We observe an opposite behavior for the hole density of the enol with an increase of the density on O1, O2, N3, N4, C7, and C8. The electron density increase not only on C5 and C6, but also on C7 and O1 (see right panel, upper row of Fig.~\ref{fig:a' and a" mo ionization}). At $t=2.8$~fs, the keto form shows a delocalization of the electron density initially localized on C6 to N3 and C8, whereas enol shows a localization of the electron density on the C6 atomic site. At $t=4.4$~fs the electron density in keto is mainly localized on C7 and N4, while for enol it appears to be delocalized on O1, C7, C8, and C6. In both cases, strong charge oscillations are anticipated immediately after the ionization, as suggested by distinct hole density at 0 fs and 1.2~fs. Nevertheless, the final distribution of the charge in both tautomers is very similar suggesting that even if the initial ultrafast dynamics differ in the first few femtoseconds, similar subsequent electron-nuclear dynamics could be expected.

Focusing now on the effect of the ionization of a' orbitals for keto, we observe an overall delocalization of the hole density with time, the main difference being the noticeable increase of the hole around H12 and C8 (see the left panel, lower row in Fig.~\ref{fig:a' and a" mo ionization}), while enol shows an opposite direction of the initial charge flow, with a decrease of hole density around C8 and H10 (see the right panel, lower row in Fig.~\ref{fig:a' and a" mo ionization}). In the a' case, the final charge distribution strongly differs between the enol and keto forms. 
 
As a consequence, it is expected that the ionization out of both a" and a' orbitals will trigger different non-adiabatic dynamics in the two forms. Allowing the charge to be delocalized all over the system, it is expected that the enol form will be more resistant to radiation damage than the keto with respect to the ionization of a' orbitals. The N4 atom in both tautomers is important to be compared, since a covalent bond with a hydrogen atom exists in the keto form and is absent in the enol one. The absence of a hydrogen atom reduces the steric constraint, making the localisation of the charge close to N4 easier. This effect is significantly stronger for the in-plane a' orbitals than for the off-plane ones with a'' symmetry, explaining the similar response of both forms in the case of the ionization out of the a'' orbitals. We can also note that the magnitude of variation of hole density per femtosecond suggests that the charge migration dynamics in enol is faster than that in keto.

\begin{figure*}
    \centering
    \includegraphics[scale = 0.35]{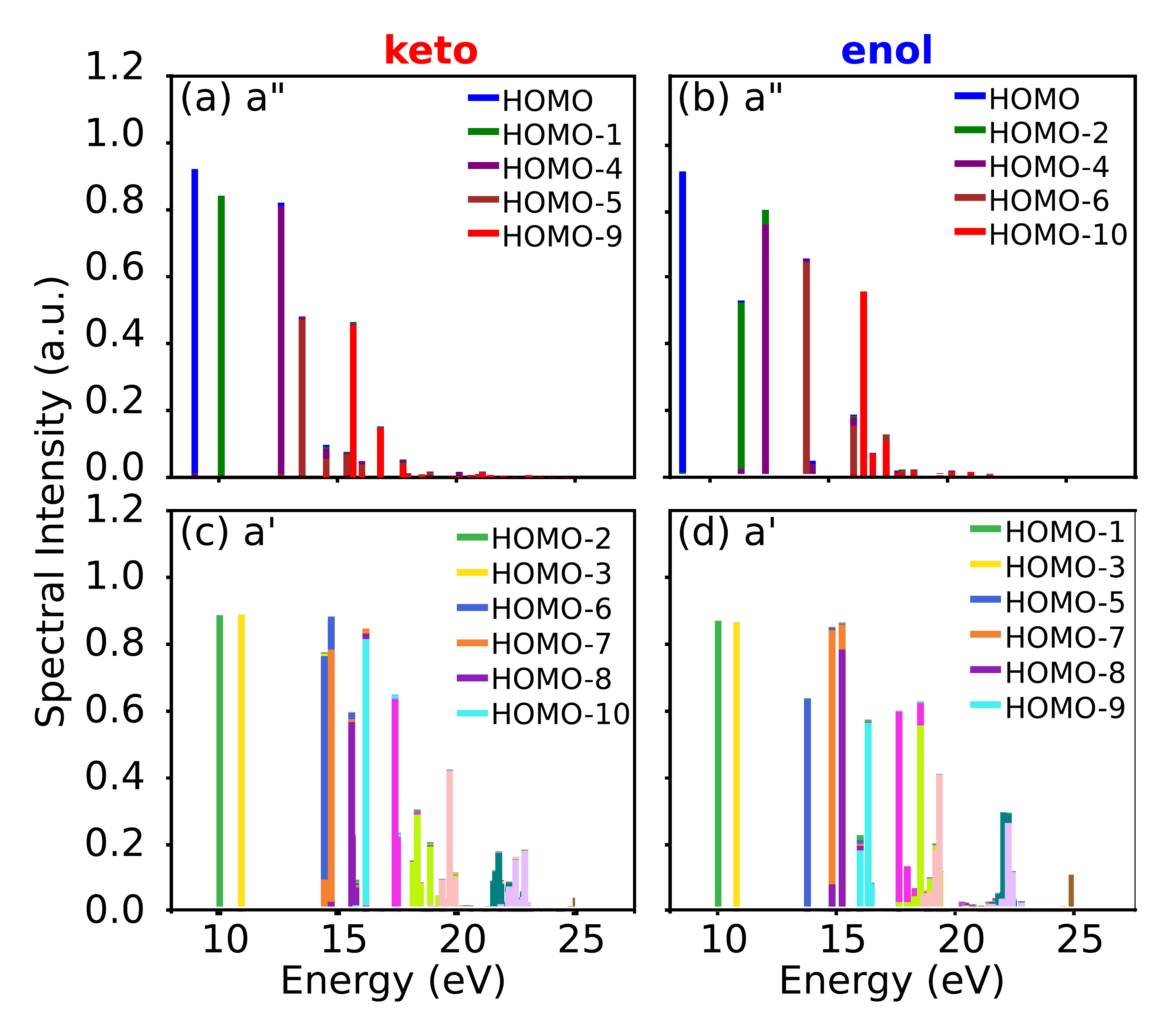}
    \caption{States with a" (upper panels) and a' (lower panels) symmetry in the ionization spectra of keto (left) and enol (right).}
    \label{fig:Uracil-K_N_E}
\end{figure*}

Let us now have a closer look on the impact of the tautomeric structures on the correlation effects. This is an important question for the possible control of charge migration for \textbf{U} using tailored IR pulses \cite{golubev2015control,golubev2017quantum}. In Fig.~\ref{fig:Uracil-K_N_E} we show only the ionic states belonging to the a'' symmetry of keto (panel (a)) and enol (panel (b)), respectively. It is apparent that many of the correlation structures are similar for keto and enol tautomers for a' as well as a'' states. Taking for example the states of a'' symmetry, it appears that the ionization of the HOMO$-5$ and HOMO$-9$ orbitals of the keto form leads to similar shake-up structures as the ionization of the HOMO$-6$ and HOMO$-10$ orbitals of the enol form. 
For a' orbital the same hole-mixing structure is apparent between the orbitals HOMO$-6$ and HOMO$-7$ of the keto form and orbitals HOMO$-7$ and HOMO$-8$ of the enol one. Hole mixing and shake-up states are the two correlation structures that one expects in this energy range. The similarities in the ionization spectra suggest that a large part of the correlation effects are conserved in tautomerization. This robustness of the correlation effects with respect to structural changes is important for the development of a molecular design for charge migration. 

It is, however, also apparent that new correlation structures appear for the enol form (compare Fig.~\ref{fig:Uracil-K_N_E}(a) and (b)). Indeed, for a'' symmetry a new hole-mixing structure exists between HOMO$-2$ and HOMO$-4$ sates that does not exist for the keto form (the respective states are more separated and the 1h configurations do not mix). The HOMO$-4$ of enol is similar to the HOMO$-4$ of keto, while the shape of HOMO$-2$ of enol differs greatly from the HOMO$-1$ of keto (see molecular orbital surface in Fig.~\ref{fig:Uracil-K_N_E_MO}). In keto, HOMO$-1$ is localized only on the O1, O2 and N3 atoms and its ionization do not lead to strong correlation effects. The HOMO$-2$ of enol is delocalized over the whole molecule leading to the hole mixing. Moreover, the main HOMO$-2$ line in enol has a large 2h1p part, while the corresponding state in keto, stemming from the ionization of HOMO$-1$ is mostly a 1h state (compare the spectral intensity of the green states in Fig.~\ref{fig:Uracil-K_N_E}(a) and (b)). New correlation structures appear also upon ionization of a' orbitals of enol. This is the case for example with the satellite structure resulting from the ionization of HOMO$-9$ that is completely missing in keto upon ionization of the corresponding HOMO$-10$ (compare the cyan states in Fig.~\ref{fig:Uracil-K_N_E}(c) and (d)). It is, however, not possible in this case to explain the difference in the spectra only by inspecting the shapes of the corresponding molecular orbitals of the two tautomers (HOMO$-9$ and HOMO$-10$, respectively), as also additional excitations are involved in forming the satellite states.

The appearance of additional and more involved correlation structures suggest that the correlation effects are somewhat stronger in the enol form. However, the global response of keto and enol structures to an XUV ionization is similar for a'' symmetry (Fig.~\ref{fig:a' and a" mo ionization}), suggesting that some correlation effects are washed-out in the averaging of the time-dependent density. Experiments able to selectively address certain energy regions in the ionization spectra or to disentangle dynamics triggered by removing an electron from individual orbitals will be necessary to unveil the richness of the correlation-driven charge migration. Further details of charge dynamics triggered by ionization of individual orbitals is discussed in the \textbf{ESI$^\dag$}. 
It is also important to note that the above analysis shows that tautomerization can both conserve and modify the correlation structure in a molecule, which makes the use of tautomeric forms a promising tool in molecular design for achieving a particular charge-migration dynamics. 

We would like now to discuss the possible experimental observation of the processes of ultrafast charge rearrangement. The schemes employed until now for studying charge migration dynamics in molecules have used both XUV and IR pulses to ionize the system and initiate the process. In the pioneering study on phenylalanine~\cite{belshaw2012,calegari2014}, the charge migration process was observed indirectly by measuring the yield of a doubly charged ionic fragment as a function of the delay between the ionizing XUV pump and the doubly ionizing IR probe pulses. An alternative technique, based on time-resolved high-harmonic generation spectroscopy~\cite{smirnova2009}, was used to reconstruct the charge migration in iodoacetylene~\cite{kraus2015measurement}. In this scheme, the molecule is field-ionized by strong IR pulses of different wavelength (800 and 1300~nm) and HHG spectrum of the rescattered electron is used as a probe. 

Although these techniques are very promising, they have certain disadvantages. Relying only on ionic fragments gives little information on the particular electron-dynamics mechanism preceding the nuclear rearrangement, while by using strong-field ionization one risks that the charge migration process is substantially modified by the IR field.
To overcome these drawbacks, here we propose an attosecond XUV-pump--XUV-probe scheme combined with Attosecond Transient Absorption Spectroscopy (ATAS) \cite{Goulielmakis2010} (see \textbf{ESI$^\dag$} Fig.~\ref{fig:experimental setup}). The attosecond duration of both the pump and the probe pulses ensures that the process of charge migration is initiated and then takes place in a field- and scattered-electron-free environment, and thus not modified by the detection scheme. At the same time, the ATAS can provide a high temporal and spatial resolution, as was recently demonstrated \cite{drescher2019,timmers2019,saito2019,zinchenko2021}.
As we discussed above, selectively following the dynamics triggered by ionization out of specific orbital or at least from orbitals belonging to particular symmetry has a lot of advantages. That is why an alignment of the molecules will be very beneficial. The latter can be obtained by an impulsive alignment of the molecular sample with a weak NIR pulse. To identify which molecular orbital was ionized by the follow-up XUV-pump pulse, we propose the use of COLd Target Recoil Ion Momentum Spectroscopy (COLTRIMS) \cite{ullrich1994cold}, where the detected photo-electrons and photo-ions are resolved both in angle and energy. Subsequently, ATAS can be used to record the temporal evolution of charge migration. Further details of our proposed schematic are in given \textbf{ESI$^\dag$}.

\section*{Conclusions}
\label{section:Conclusion}

In summary, we studied the ultrafast charge migration dynamics between different atomic sites, occurring after an XUV photo-ionization of keto and enol tautomeric forms of a prototype molecule \textbf{U}. We reported and compared ionization spectra of keto and enol forms, computed with high-level \textit{ab initio} approach, the Green's function ADC(3) method. The spectra show that although the first few cationic states are nearly pure 1h states, a signature for weak electron-correlation effects, states beyond 12~eV show strong correlation and thus ionization out of the corresponding lower-lying valence orbitals sets in a prominent ultrafast inter-site electron dynamics. To better understand the XUV ionization processes appearing in nature, we studied the overall electron dynamics triggered by the ionization of all outer-valence orbitals (7 in total) populating states up to about 15~eV in the ionization spectra, and compared the ultrafast response in the two tautomers. We showed that even though such averaging washes out many of the correlation-driven charge-migration dynamics that take place, the results clearly show that the different tautomers respond differently to an XUV ionization. Indeed, the charge migration is noticeably different, with different patterns and timescales, and with a charge flow going in different directions. To detect and reconstruct such ultrafast charge dynamics experimentally, we also propose an experimental schematics. 

The ultrafast charge migration will, of course, influence and to some extent predetermine the slower nuclear dynamics. Therefore, the significantly different charge distributions in the two tautomeric forms of \textbf{U} resulting from the ultrafast electron dynamics can be expected to lead to different fragmentation channels. Different fragmentation patterns, or different charge-directed reactivity \cite{remacle1998charge}, was observed in peptide chains and stereoisomers of amides in a series of pioneering experiments performed by Schlag, Weinkauf, and Müller-Dethlefs~\cite{weinkauf1995elementary,weinkauf2008ultrafast,ullrich2001zeke}. These experiments showed that the charge migration distance, as well as the fragmentation pattern are very sensitive to the particular amino-acid sequence in the peptide chains and conformation of amide bonds. Our results now show that even the different tautomers can exhibit different charge-migration dynamics. Thus, our study may bring a key understanding of why different tautomers sometimes show prominently different fragmentation patterns. Additionally, one of the key obstacles in the experimental study of charge migration is to find suitable molecules. In this respect, our study provides important information that can drive future experiments. Our results also suggest that using tautomeric forms, which may exhibit prominently different electron dynamics after ionization, may be a promising path in the development of molecular design for desired charge migration.  We hope that our findings will motivate further theoretical and experimental work in this direction.

\section*{Acknowledgements}
ELI-ALPS is supported by the European Union and co-financed by the European Regional Development Fund (GINOP-2.3.6-15-2015-00001). KC and MUK acknowledge Project no. 2019-2.1.13-TÉT-IN-2020-00059 which has been implemented with the support provided from the National Research, Development and Innovation Fund of Hungary, financed under the 2019-2.1.13-TÉT-IN funding scheme. KC and MUK also acknowledges funding from PaNOSC European project. VD and AIK acknowledge financial support from the DFG through the QUTIF priority programme.

\bibliography{uracil}

\clearpage
\newpage
\Large{\textbf{Supplementary Information}\setcounter{section}{0}}
\small
\setcounter{figure}{0}
\setcounter{table}{0}
\renewcommand{\theequation}{S\arabic{equation}}
\renewcommand{\thefigure}{S\arabic{figure}}
\renewcommand{\thetable}{S\arabic{table}}

\section*{Computed bond lengths and Molecular Orbitals}

The geometry optimization for the keto and enol tautomers was done at PBE0/def2-TZVP level. The resulting bond lengths, together with known experimental results for keto \textbf{U} are listed in \textbf{ESI$^\dag$} Table.~\ref{tab:Bond lengths}. The computed bond length values for keto show good match with the experimental results~\cite{harsanyi1986equilibrium}.

\begin{table}[H]
\centering
\begin{tabular}{|c|c|c|}
\hline
    bonds & keto (Ref.~\cite{harsanyi1986equilibrium}) & enol \\
    \hline
  C5-O1   & 1.2074 (1.212) & 1.3297 \\
  C6-O2    & 1.2096 (1.212) & 1.2112  \\
  C5-N3     & 1.372 (1.391) & 1.3427 \\
  C6-N3  & 1.3992 (1.415) & 1.4102 \\
  C5-N4   & 1.3823 (1.395) & 1.2919 \\
  C6-C8    & 1.4514 (1.462) & 1.440 \\
  C7-C8  & 1.3416 (1.343) & 1.3547 \\
  C7-N4 & 1.364 (1.396) & 1.3639 \\ \hline
\end{tabular}
\caption{Calculated bond lengths (in \AA) for keto (experimental data~\cite{harsanyi1986equilibrium}) and enol tautomers.}
\label{tab:Bond lengths}
\end{table}

\begin{figure*}
    \centering
    \includegraphics[scale = 0.80]{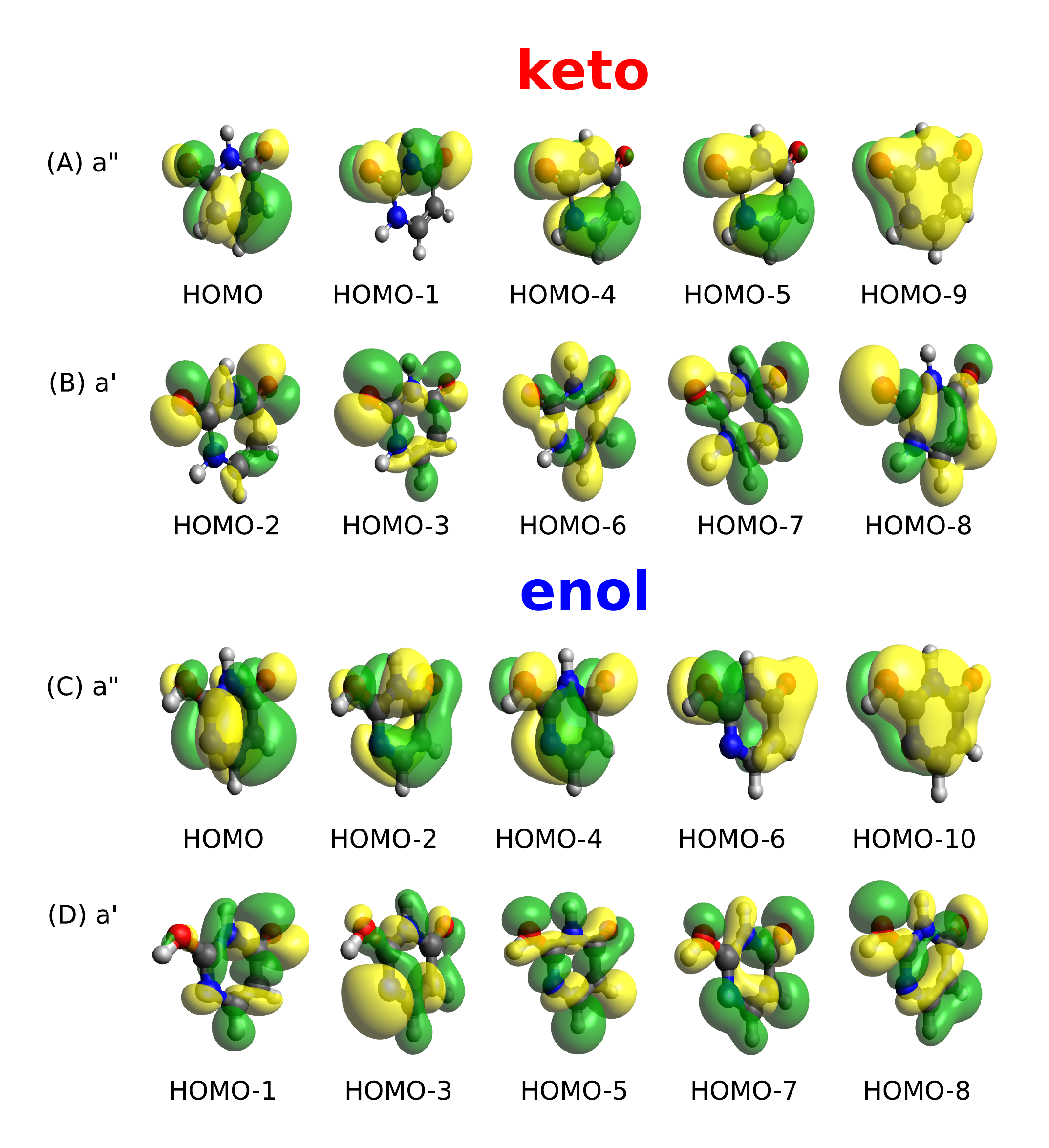}
    \caption{The 5 major orbitals from the a" (A, C) and a' (B, D) symmetry in the ionization spectra of keto (upper two panels) and enol (lower two panels) two symmetries are shown with isosurfaces of 0.02 e/\AA$^3$.}
    \label{fig:Uracil-K_N_E_MO}
\end{figure*}

\section*{Dynamics initiated by ionization of individual orbitals}

Interesting charge migration dynamics after individual orbital ionization was observed for HOMO$-4$, HOMO$-5$, and HOMO$-9$ molecular orbitals of keto and HOMO$-1$, HOMO$-4$, HOMO$-6$ molecular orbitals of enol. These charge migration dynamics will be presented in detail in the following subsections.

\subsection*{Ionization of HOMO$-4$ of keto tautomer}
The ionization of the HOMO$-4$ of the keto tautomer leads to a typical satellite structure, the specificity being that the satellite posses 1h configuration from both HOMO$-4$ and HOMO$-5$ (see Fig.~\ref{fig:Uracil-K_N_E}(a) in main text). As a result, the ionization of HOMO$-4$ will coherently populate two cationic states. This will lead to oscillation of residual charge with a period of $t = \frac{2\pi}{\omega}$ with $\omega = \frac{E_{S}-E_{M}}{\hbar}$, where $E_{M}$ is the ionization energy of the main state and $E_{S}$ the one of the satellite. We observe at each quarter cycle an avoided crossing as shown in \textbf{ESI$^\dag$} Fig.~\ref{fig:ofile_merged_uk_ue}(a) where the two natural charge orbitals interchange their character between HOMO$-4$ and HOMO$-5$ (shown by change in colours along the curves for HOMO$-4$ and HOMO$-5$). Directly after the sudden ionization of HOMO$-4$ we see an increase in charge density on O1 (circled part on the molecular structure (shown in \textbf{ESI$^\dag$} Fig.~\ref{fig:ofile_merged_uk_ue}(a)). Then we observe the increase in hole occupancy on HOMO$-5$ and decrease on HOMO$-4$ leading to the increase of the charge density on N4 and its decrease on O1, respectively. At the crossing points (observed at 0.88~fs and 3.12~fs), we see both orbitals to contribute to charge density on O1 and N4. Whereas, the charge density on O1 and N4 originates mainly from HOMO$-5$ (at 2.26~fs) and HOMO$-4$ (at 4.15~fs).

\subsection*{Ionization of HOMO$-5$ of keto tautomer}
\begin{figure*}
\centering
\includegraphics[scale = 0.53]{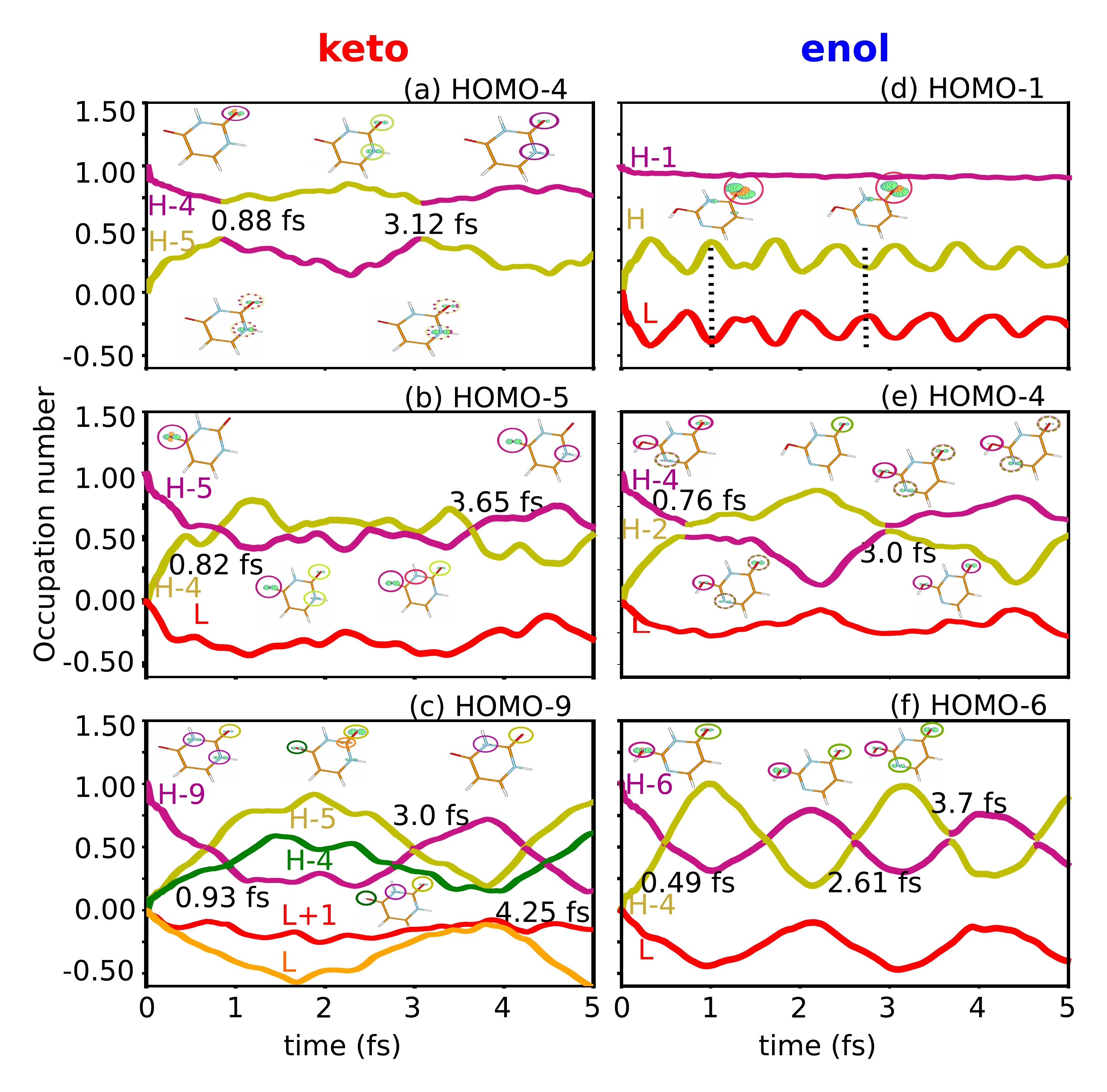}
\caption{Time evolution of the leading occupation number $\tilde n_p(t)$ (see Eq.~(\ref{eq:q(r,t)} in main text)) after ionization of a molecular orbital of uracil tautomer. Snapshots of the charge density are also shown. The colours of the circles around the charge distribution on the molecular structures show the corresponding molecular orbitals contributing to the charge density at each position. (a) HOMO$-4$, (b) HOMO$-5$, (c) HOMO$-9$ of keto tautomer (iso-value 0.01 e/\AA$^{3}$, hole density in green and electron density in orange) and (d) HOMO$-1$, (e) HOMO$-4$, (f) HOMO$-6$ of enol tautomer of uracil (with an iso-value of 0.03 charge/\AA$^{3}$). }
\label{fig:ofile_merged_uk_ue}
\end{figure*}

From the ionization spectra presented in main text Fig.~\ref{fig:Uracil-K_N_E}(a) we observe satellites structures populated by the ionization of HOMO$-5$ with a main state at 13.5~eV and satellite states at 14.43~eV and 15.39~eV. We see a significant contribution from 1h configuration of both HOMO$-5$ and HOMO$-4$ in the satellite at 13.5~eV. The dynamics resulting from the ionization of HOMO$-5$ is presented in \textbf{ESI$^\dag$} Fig.~\ref{fig:ofile_merged_uk_ue}(b).
We see a prominent interchange of the molecular orbital character between HOMO$-5$ and HOMO$-4$ at the crossing points (observed at 0.82~fs and 3.65~fs). After the ionization, the hole density on O2 first increases and is then distributed on O1 and N4 as seen at 0.82~fs. To better understands the contribution of the involved molecular orbitals, we analysed the charge density at two other time steps, 3.39~fs (before the crossing at 3.65~fs) and 4.5~fs (after the crossing) where the contribution from HOMO$-4$ and HOMO$-5$ is high respectively. We found that at 3.39~fs the charge density contribution on O1 comes from HOMO$-4$, on N3 from LUMO, and on N4 and O2 from HOMO$-5$. Hence, due to correlation effects the charge density propagates with time from O2 to N4.

\subsection*{Ionization of HOMO$-9$ of keto tautomer}
The ionization of HOMO$-9$ leads to population of several states due to the presence of a complex satellite structure (see Fig.~\ref{fig:Uracil-K_N_E}(a) in main text). After the ionization, we observe that hole density is created on N4, O1, and N3. The 2h1p state is dominated by two different configurations; one with holes in HOMO$-9$ and HOMO$-5$ and an excited electron to LUMO, and another one with holes in HOMO$-9$ and HOMO$-4$ and a particle in LUMO$+1$. The charge migration dynamics is presented \textbf{ESI$^\dag$} Fig.~\ref{fig:ofile_merged_uk_ue}(c). The initially created hole density on N4 decreases at the first crossing point and at 1.88~fs we observe an increased hole density on O1, O2, and N3 with an electron density on C5. Therefore, we observe here a prominent charge migration from O1, N3 and N4 to O1, O2 and C5.

\subsection*{Ionization of HOMO$-1$ of enol tautomer}
Ionization of HOMO$-1$ leads to the population of a main state at 10.02~eV and a relaxation satellite at 16.0~eV (see Fig.~\ref{fig:Uracil-K_N_E}(d) in main text). The principal 2h1p configuration of the relaxation satellite is holes in HOMO and HOMO$-1$ and a particle in LUMO. In this case the initially created hole stays nearly constant, but an excitation-deexcitation dynamics is triggered between HOMO and LUMO (see \textbf{ESI$^\dag$} Fig.~\ref{fig:ofile_merged_uk_ue}(d)). Immediately after the ionization, at 0.04~fs, the charge density is mainly localized on the oxygen atom O2. We observe small oscillations of the charge density between O2, and N3 and C8. The charge density on O2 and N3 is more pronounced than on C8 and is thus observed clearly through density plots (with an iso-value of 0.03 charge/\AA$^{3}$).

\subsection*{Ionization of HOMO$-4$ of enol tautomer}
The ionization of HOMO$-4$ leads to a complex satellites structure with in particular hole mixing character with HOMO$-2$ for states at 10.99~eV and 12.01~eV (see Fig.~\ref{fig:Uracil-K_N_E}(c) in main text). The ionization of HOMO$-4$ creates hole density on O1, O2, and N4 (see \textbf{ESI$^\dag$} Fig.~\ref{fig:ofile_merged_uk_ue}(e)). After 0.76~fs, about 40\% of the initial charge is redistributed between HOMO$-4$ and HOMO$-2$ orbitals, accompanied by an additional excitation of a fraction of an electron (0.28) to LUMO. After 0.76~fs, we observe an avoided crossing and the character of the corresponding natural charge orbitals is interchanged, i.e. HOMO$-4$ continues to lose occupation (the hole gets filled) and HOMO$-2$ to gain occupation (the hole continues to open). This process lasts until 2.17~fs, when more than 88\% of initial hole charge has migrated to HOMO$-2$. This exchange of charge density keeps repeating with a period of about 4.4~fs. Looking at the charge density plots, at the 0.76~fs and 3.06~fs crossing points, we see that the hole density is mainly on O1, O2, and N4 with some electron density on O1. The excess electron density is due to the increase in occupation number of LUMO resulting from the accompanying excitation. At the peak of the process (at 2.17~fs), the positive charge is mainly on O2. Therefore, the process observed after ionization of HOMO$-4$ is a charge migration from O1 and N4 to O2.

\subsection*{Ionization of HOMO$-6$ of enol tautomer}
The ionization of HOMO$-6$ leads to a complex satellite structure with a main state at 13.74~eV and satellite states at 13.99~eV, 15.72~eV and 17.1~eV. The cationic states at 13.99~eV and 15.72~eV show a small 1h contribution from HOMO$-4$ (see Fig.~\ref{fig:Uracil-K_N_E}(c) in main text). The evolution of the main hole occupation numbers after ionization of HOMO$-6$ of enol is shown in \textbf{ESI$^\dag$} Fig.~\ref{fig:ofile_merged_uk_ue}(f). Immediately after the ionization of HOMO$-6$, hole density is localized on O1 and O2. As time proceeds, the hole migrates to HOMO$-4$ which results in formation of a hole density on N4. The process is accompanied by excitation for an electron to LUMO. The overall process represents oscillations of hole density between O2 and O1, as well as oscillations of charge density on O2 and N3, where hole density on O2 and electron density on N3 increase and decrease periodically. 

In summary, photo-ionization of individual orbitals of both the tautomers show different behaviour with different patters and timescales. We observe, clear charge oscillation between N4 and C5, with period of 1.88 fs on ionization of HOMO-9 of keto, and, relatively stronger and faster beating between N4, C5 and O1 with period of 0.98 fs on ionization of HOMO-6 of enol tautomer. This shows the enol leads to stronger correlation effects and fast charge migration compared to keto. 

\begin{figure*}
    \centering
    \includegraphics[scale=0.8]{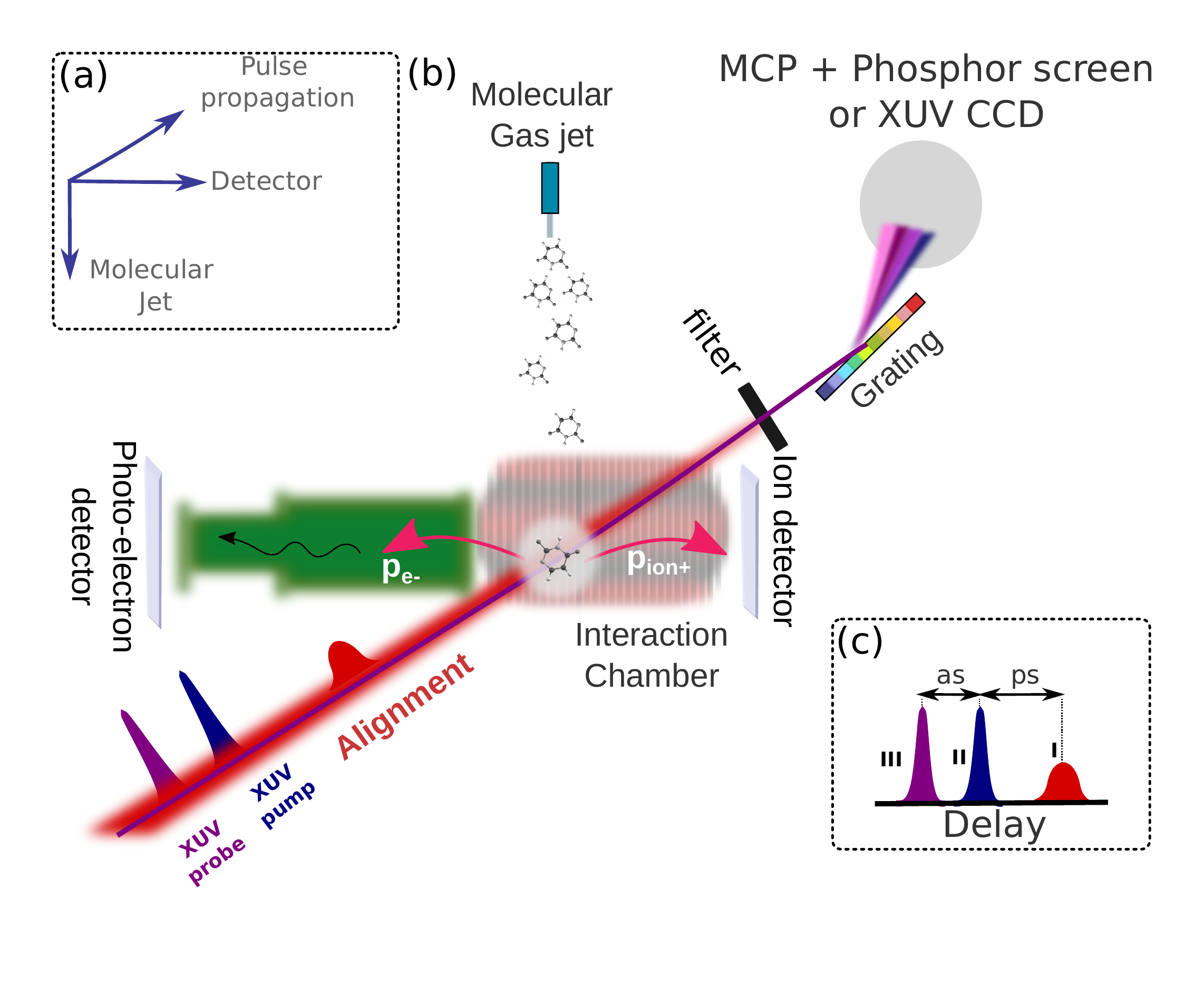}
    \caption{Schematics for the proposed experimental setup with (a) showing the direction of the molecular jet, pulse propagation and detectors. (b) The experiment is executed in two independent steps: (i) The molecular gas jet with molecules entering the interaction chamber are aligned with a long NIR alignment pulse. The aligned molecules are ionized with \textit{as} XUV-pump pulse and the resulting photoelectrons and ions are detected by the COLTRIMS to identify the orbitals involved in photo-ionization. (ii) A delayed \textit{as} XUV-probe pulse interacts with the target and captures the temporal evolution of the charge migration.  The later is observed with ATAS where the XUV pulse after reflection from grating is captured by an XUV photon-detector (assembled with a micro-channel plate (MCP) and phosphor screen or XUV CCD). The alignment NIR-pulse is blocked by a filter placed between the gas target and the grating. (c) Shows the time delay between the three pulses (alignment, XUV-pump and the XUV-probe) when entering the interaction chamber.}
    \label{fig:experimental setup}
\end{figure*}

\section*{Schematics of the proposed experimental setup}
\label{SI sect: Scehmatics}

Several pioneering experiments have been performed to address charge dynamics \cite{stener2012recoil, guillemin2015selecting, teng2018attosecond}, yet given the complexity of the process it is difficult to extract all information at once. To study the electron dynamics during the first few femtoseconds after the ionization, we propose to use the experimental schematics shown in Fig~\ref{fig:experimental setup}. The three major steps envisaged are: (i) alignment of the molecules using non-ionizing NIR pulse with duration of $\sim$100~fs, (ii) ionization of the molecule with an \textit{as} XUV-pump pulse, and (iii) tracing the charge migration dynamics in the ion using a second XUV-probe pulse by measuring its absorption as a function of the pump-probe delay. 

Impulsive alignment of the molecule can be achieved using long pulse (for example THz or NIR pulse) \cite{seideman1995rotational}. Note that controlling the alignment of large molecules is certainly challenging. However, the topic of aligning asymmetric-top molecules in 3D in a field-free environment is developing fast. For the molecule under consideration, 3D alignment can be achieved either through picosecond pulse shaping~\cite{mullins2020picosecond} or by using multipulse techniques. In either case, following the metric for 3D alignment~\cite{ren2014multipulse} a typical alignment of $\langle \cos^2(\delta) \rangle \sim 0.8$ can be expected, as TDSE simulations on Uracil have shown (see figure 3.16 in ref.~\cite{makhija2014laser}). This will be sufficient for a potential experiment, in which the subsequent charge dynamics signatures will be captured in the Recoil-ion and electron momentum with COLTRIMS~\cite{ullrich2003recoil}.

After alignment of molecules, the ionization of the valence molecular orbitals (MO) can by be performed with a linearly polarized attosecond XUV pulse. In such a way, we will be able to selectively address the orbitals belonging to a' or a" symmetries. We can further extract the contribution of different MOs with the help of COLTRIMS techniques \cite{ullrich1994cold}. The photoelectron emission direction is related to the shape of the MO which is ionized, and thus angular resolved photoelectron-photoion coincidence map may allow to identify the dominant MOs bearing the initial hole. 

The hole migration triggered after the ionization of \textbf{U} can be captured with the help of attosecond transient absorption spectroscopy (ATAS). A delayed second \textit{as} XUV-pulse can probe the charge dynamics, by driving resonant transitions in the ion which will depend on the momentary charge distribution in the molecule. In such a way, at each time delay, the absorption/transmission spectrogram will take snapshot of the actual electron distribution in the molecule. The transmitted XUV pulse can be detected with an XUV spectrometer (grating, micro-channel plate (MCP) and phosphor screen or XUV CCD), as shown in \textbf{ESI$^\dag$} Fig.~\ref{fig:experimental setup}. 

The proposed experiment is technically very challenging. The state-of-the-art HHG-driven \textit{as} XUV beamlines are yet to overcome the bottlenecks of implementing XUV-pump--XUV-probe experiments. However, there are a few research facilities, like ELI-ALPS, that hold the prospects to host such experiments in the near future. 
Furthermore, the proposed experiment relies on aligned molecules which in combination with ATAS has some limitations. As temperature of the molecules have a large impact on the degree of molecular alignment, one usually uses supersonic jets for preparing the sample, which however have relatively low density of molecules. ATAS, on the other side, relies on high density samples for achieving a good signal to noise ration. We will, therefore, probably need to make a compromise between the degree of alignment and the molecular beam density, such that we still get a good alignment, and thus selectivity, and perform ATAS \cite{peng2019symmetry}. This can be solved by performing the COLTRIMS measurement first with a lower density gas jet, then after switching off the sensitive electronics one can increase the pressure which makes absorption of the XUV measureable. 
\end{document}